\documentclass[twoside,a4paper,11pt]{sea10}
\usepackage{graphicx}
\usepackage{hyperref}
\usepackage{movie15} 
\def\xmm{{\it XMM--Newton}}
\def\mnras{MNRAS}       
\def\apj{ApJ}           
\def\apjs{ApJS}         
\def\aap{A\&A}          
\def\apjl{ApJ Letters}  
\newcommand{\xr}{X--ray}
\newcommand{\kms}{km\,s$^{-1}$}
\newcommand{\ie}{i.e.}
\newcommand{\kev}{keV}
\newcommand{\ergcms}{erg\,cm$^{-2}$\,s$^{-1}$}
\newcommand{\swi}{SWIFT\,J2127.4+5654}
\begin{document}
\pagenumbering{arabic}
\pagestyle{myheadings}
\thispagestyle{empty}
{\flushleft\includegraphics[width=\textwidth,bb=58 650 590 680]{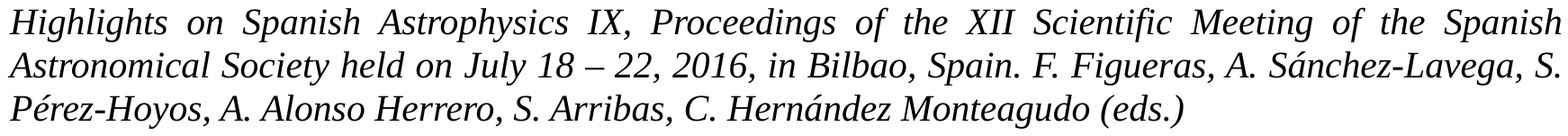}}
\vspace*{0.2cm}
\begin{flushleft}
{\bf {\LARGE
%
Probing relativistic effects in the central engine of AGN
}\\
\vspace*{1cm}
%
Mario Sanfrutos$^{1}$ 
and
Giovanni Miniutti$^{1}$
%
}\\
\vspace*{0.5cm}
%
$^{1}$
Centro de Astrobiolog\'{i}a (CSIC--INTA), Dep. de Astrof\'{i}sica;
ESAC, Villanueva de la Ca\~nada, E-28692 Madrid, Spain
%
\end{flushleft}
%
\markboth{
Probing relativistic effects in the central engine of AGN%
}{ 
%
M. Sanfrutos \& G. Miniutti
%
}
\thispagestyle{empty}
\vspace*{0.4cm}
\begin{minipage}[l]{0.09\textwidth}
\ 
\end{minipage}
\begin{minipage}[r]{0.9\textwidth}
\vspace{1cm}
\section*{Abstract}{\small
%
Active Galactic Nuclei (AGN) are perfect laboratories to check General Relativity (GR) effects by using Broad Line Region (BLR) clouds eclipses to probe the innermost regions of the accretion disk. 
A new relativistic \xr\ spectral model for \xr\ eclipses is introduced. First we present the different observables that are involved in \xr\ eclipses, including the \xr\ emitting regions size, the emissivity index, the cloud's column density, ionization, size and velocity, the black hole spin, and the system's inclination. 
Then we highlight some theoretical predictions on the observables by using \xmm\ simulations, finding that  absorption varies depending on the photons' energy range, being maximum when the approaching side of the \xr --emitting region is covered. 
Finally, we fit our relativistic model to actual \xmm\ data from a long observation of the NLS1 galaxy \swi , and compare our results with a previous work, in which we addressed the BLR cloud eclipse from a non--relativistic prespective. 

%
\normalsize}
\end{minipage}
%
%
%
\section{Introduction \label{intro}}

The innermost regions in AGN are in a strong gravity regime due to their close proximities to the central supermassive black hole (SMBH). 
Hence, the \xr\ emission reprocessed by the inner accretion disk is imprinted by GR effects such as Doppler boosting, gravitational redshift and light bending, that shape the spectral features including the K$\alpha$ emission line profile \cite{Reynolds03}. 
\xr\ time-resolved spectroscopy of AGN is therefore our ultimate tool to probe GR effects under extreme gravity conditions. 
The emission from the approaching side of the disk is enhanced due to Doppler boosting, while it is diminished from the receding side. 
Since the central engine is unresolved, obscuration by optically thick matter of the different \xr\ emitting regions by structures in our line of sight (LOS) is a promising manner to distinguish the reflection from different parts of the disk \cite{mckernan98}, and a method to investigate relativistic effects in AGN \xr\ spectra by means of eclipses has been proposed \cite{risaliti2011relat}. 

Variable \xr\ absorption in AGN has been noticed on all time scales, not depending of their luminosity or morphology \cite{marshall81, turner99_var}. The absorber has been identified with clouds of the dusty, clumpy torus at the pc--scales and long timescales \cite{marinucci2013, risaliti_var2002}, or the BLR at short timescales \cite{bianchi2009, elvis2004, maiolino10, puccetti2007, Risaliti09, Risaliti11_mrk766, sanfrutos2013}. In the latter, the observed absorption variability points to cloud sizes of the order of few gravitational radii, thus comparable to the \xr\ emitting regions. So, detailed modeling of such events can enable us to draw an accurate picture of the system's geometry. This high variability has been explained through fast column density changes as a result of material in the BLR crossing our LOS \cite{miniutti2014, sanfrutos2016a}.

\section{The relativistic model \label{model}}

AGN spectra are composed of at least two components: the continuum power law emission, from the corona, and the reflection--dominated component, from the accretion disk \cite{fabian_miniutti04, fabian_miniutti05, Ponti06, Vaughan04}. Both components arise from regions of the accretion flow only a few~$r_{\rm g}$ away from the central SMBH \cite{Ponti06, iwasawa97}. 
The geometry that we assume for the system consists of a SMBH, characterized by its mass ($M_{\rm BH}$) and spin parameter (${\rm a^*}$), around which an optically thick but physically thin ionized accretion disk extends from the innermost stable circular orbit (ISCO) out to $400\,r_{\rm g}$. 
Several configurations have been proposed for the disk and the continuum \xr\ source: 
the hot and radiatively compact spherical corona with radius between $3$ and $10\,r_{\rm g}$ \cite{Fabian15}; 
the slab geometry \cite{rozanska15};  
the patchy structure \cite{stern95, Wilkins15a}; 
and the jet base interpretation \cite{Wilkins15b}, 
just to mention some of them. 
We adhere to the slab geometry: in the following, the corona is a parallel plane at a negligible distance above the disk. The inner radius of the corona is forced to be coincidental with the ISCO, while its outer radius is only a few $r_{\rm g}$. 
For the reprocessed component, we assume it to arise from the innermost regions of the accretion disk, between the ISCO and an outer limit determined by the disk's emissivity index which is constrained to satisfy that at least $99\%$ of the reprocessed emission comes from the innermost $\sim$\,$12$\,$r_{\rm g}$, \ie\ $q \geq 3$. 
The eclipsing cloud is assumed to be co--rotating with the disk at a velocity of 3000\,\kms , typical of the BLR of the system. 
Since these kind of obscuration events are more likely to be detected in unobscured sources, we disregard the torus of the Unification models within the framework of this work. 

We build our model within {\small{XSPEC}} \cite{arnaud_xspec1996}, 
as a power law accounting for the \xr\ continuum, 
plus the \xr\ reflection code {\small XILLVER} \cite{Garcia13} accounting for the reflection component. 
These two components are multiplied by the {\small KYNCONV} relativistic convolution model \cite{dovciak2004}, 
which adds all relativistic effects due to strong gravity and fast motions close to the SMBH, 
allowing to obscure part of the emission with a circular cloud whose size and position can be determined. 
We model the accretion disk to cover at least one half of the sky as seen from the central engine, with a typical ionization of $\xi^{\rm (disk)} \sim 30$\,erg\,cm\,s$^{-1}$ and solar abundances. The energy cutoff is frozen to 300\,\kev . 
The main parameters that can be tuned are: 
the inclination $\theta$ of the system with respect to our LOS; 
the black hole spin ${\rm a}^*$; 
the emissivity index ${\rm q}$; 
the cloud position; 
and the cloud size. 
The column density $N_{\rm H}^{\rm (cl.)}$ and ionization log\,$\xi^{\rm (cl.)}$ of the cloud are set by the ionized absorption code {\small{ZXIPCF}} \cite{reeves_zxipcf2008}. 
We model the Galactic absorption as a neutral absorption component. 
We build this model as an upgrade of our previous GR-model \cite{sanfrutos2016b}. 
Below we explain the role of every parameter involved in it. 

Absorption due to material in our Galaxy is modelled by means of the photoelectric absorption component {\small{PHABS}} in {\small{XSPEC}}. We choose an intermediate arbitrary intermediate value of $N_{\rm H}^{\rm (Gal.)} = 3 \times 10^{20}\,{\rm cm}^{-2}$. Absorption due to partially ionized material in the host galaxy is modelled with {\small{ZXIPCF}}. The equivalent column density ranges from $10^{23}\,{\rm cm}^{-2}$ (Compton--thin) to $5\times 10^{24}\,{\rm cm}^{-2}$ (Compton--thick), typical for clouds in the BLR. Its ionization takes values between $-3$ (neutral) and $2$ (highly ionized material). The radius and position of the cloud are determined by a set of parameters of the relativistic code {\small{KYNCONV}}, all in terms of gravitational radii. 
The primary continuum emission from the corona is modelled with a simple power law component with photon index set to a typical value of $\Gamma = 2$. The continuum normalization is chosen so that the unabsorbed $2-10$\,\kev\ \xr\ flux is $10^{-11}$\,\ergcms , typical for bright, local AGN. 
An exponential high--energy cutoff of $E_c = 300$\,\kev\ is set. 
The reprocessed emission is modelled by means of the \xr\ reflection code {\small{XILLVER}} \cite{Garcia13}. The photon index is the same as for the continuum. The iron abundance is set to solar and the disk ionization is fixed to a typical value of log\,$\xi^{\rm (disk)} = 1.5$ in order to allow the existence of enough reflection features. 
We set the same cutoff as for the power law. 
Finally, the reflection normalization is fixed in order to get a reflection fraction between 1 and 2, \ie\ the disk to cover at least one half of the sky as seen from the central engine, in agreement with observations. 

\begin{figure}
\begin{center}
\includegraphics[width=0.618\textwidth]{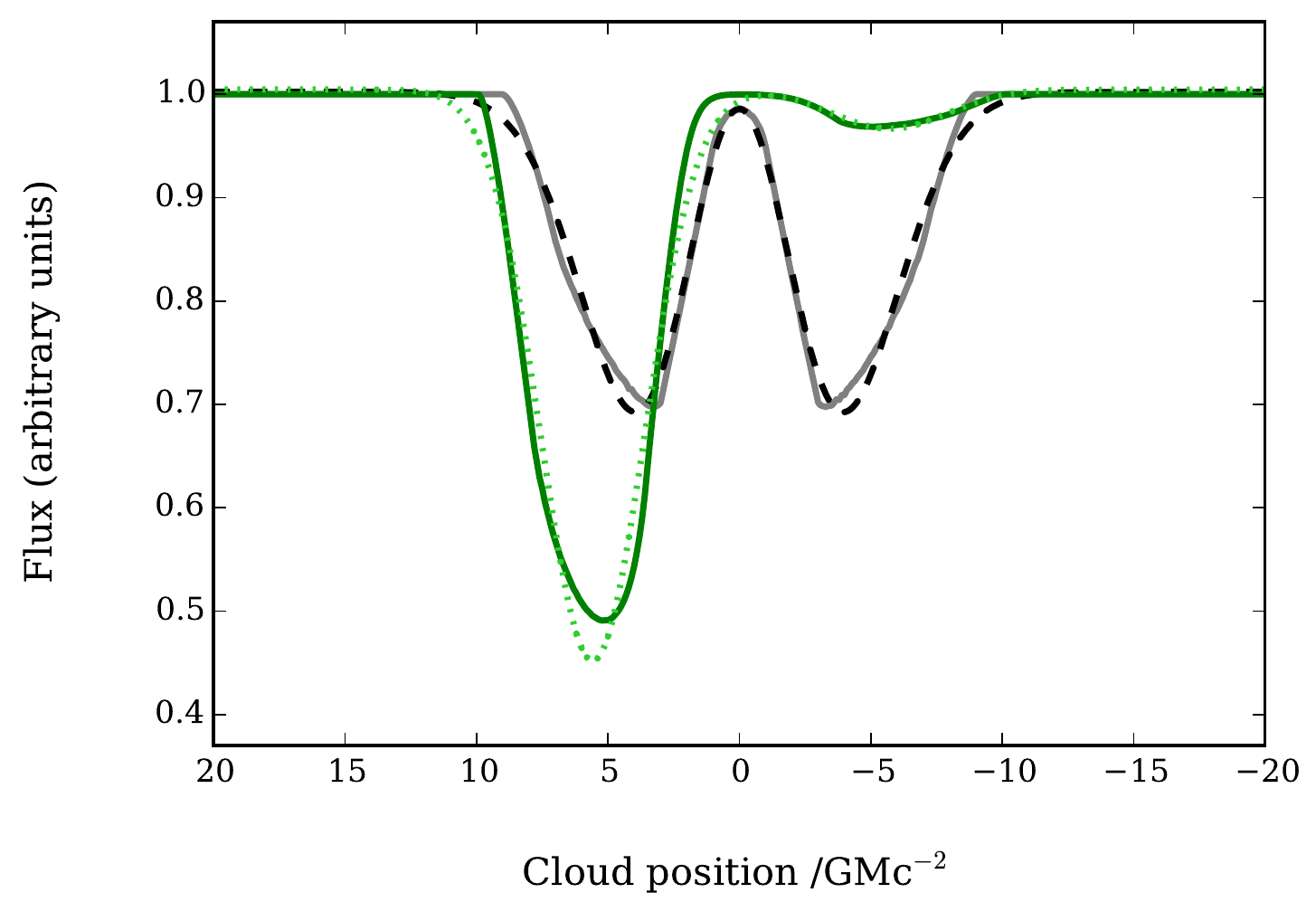}
\caption{\label{comparison_newton_rel} Lightcurves created by a circular cloud eclipsing a plane--parallel annular source inclined $45^{\circ}$. The cloud radius is $R_{\rm c} = 3$\,$r_{\rm g}$ and the annular source has inner and outer radii of $R_{\rm s}^{\rm in} = 4$\,$r_{\rm g}$ and $R_{\rm s}^{\rm out} = 6$\,$r_{\rm g}$ respectively. 
The solid grey line is the Monte Carlo simulated flux during an eclipse involving a Newtonian isotropically--emitting disk, and the dashed black line represents the fit to two gaussians. 
The solid green line is the flux simulated from the {\small KYNCONV} relativistic model, emitted from a disk around a rotating black hole with spin $a = 0.56$ (ISCO = $4$\,$r_{\rm g}$). The dotted green line is the fit to two gaussians. 
}
\end{center}
\end{figure}

The inclination of the system, defined as the angle between the normal to the disk and our LOS, directly determines how asymmetric the flux-lightcurves are during an eclipse with respect to the position of the black hole. When the system is observed face--on, Doppler boosting is undetectable from our point of view: there are no approaching nor receding regions in the disk, since its orbital plane is perpendicular to our LOS. Therefore the flux-profiles observed are symmetric. The larger the inclination is, the greater is the Doppler boosting effect, so that the more noticeable is the asymmetry of these profiles, independently of every other parameter. An example of a lightcurve produced during an eclipse of a plane--parallel annular \xr\ source inclined $45^{\circ}$ is shown in Fig.~\ref{comparison_newton_rel}. The Newtonian case is included in the figure in order to ease comparison with the GR one. Notice that the flux from the approaching side of the disk (left wing) is several times greater than that from the receding side (right wing) in the GR-dominated scenario. 

Different values of the ISCO, and hence the SMBH spin, imprint the eclipse lightcurves too. In the cases where the cloud size is comparable to the ISCO, it is possible to measure the ISCO with great precision in an unprecedented manner when the size of the cloud is known, by simply measuring the distance between the peak preceding and following the SMBH position in the lightcurves. When larger clouds are involved in the eclipse, GR effects are still detectable, although measurement of the ISCO becomes infeasible. 

It is important to mention that the secondary peaks of the lightcurves obtained in some of the configurations described here are beyond the limits of detectability of available instruments. However, detectability by future \xr\ observatories such as the ATHENA mission must be further investigated. Yet, most of the actual physical situations still represent interesting cases that we can discern by using current technology.

\section{Spectral--fitting to \swi\ data}
\label{sec:eclipses:spectral_fitting}

The singular shapes of the light curves described in Section\,\ref{model} can explain the second rise observed in some other light curves detected in actual observations taken during BLR clouds eclipses. Variable partial covering in the time--resolved spectral analysis of the Seyfert galaxy NGC\,1365 is well explained by a model consisting of a constant continuum absorbed by a constant column density, plus a variable partial--covering component \cite{Risaliti09}. The covering fraction variations are highly significant and clearly show the signature of a complete transit of an eclipsing cloud. 
The same behaviour is shown by \swi\ \cite{sanfrutos2013}. This inspires us to dig into the analysis of these relativistic effects by means of fits to real spectra. In the following paragraphs, we describe our study of GR effects in the eclipse detected in our \xmm\ observation of \swi . 

First we fix some parameters based in previous unabsorbed observations. 
The inclination angle is fixed to the common value of $i=45^\circ$ \cite{miniutti2009, Patrick11}. Also, we fix the SMBH spin to $a = 0.58$ \cite{marinucci2014, miniutti2009, Patrick11, sanfrutos2013}. 
The photon index is fixed to $\Gamma = 2$ \cite{sanfrutos2013}. 
The emissivity index of the disk is fixed to $q = 4.9$ \cite{sanfrutos2013}. 
Metal abundances are consistently set to solar \cite{marinucci2014, miniutti2009, Patrick11}. The ionization of the reflector is fixed to $\xi = 10$\,erg\,cm\,s$^{-1}$ \cite{sanfrutos2013}. We adopt the high--energy cutoff $E_c = 108$\,\kev\ \cite{marinucci2014}. 
We fix the column density of the neutral eclipsing cloud to $N_{\rm H} = 2\times10^{22}$\,cm$^{-2}$ \cite{sanfrutos2013}. 
Finally, we assume that the cloud follows a Keplerian orbit co--rotating with the SMBH and the accretion disk, moving at velocity $v_c \leq 2100$\,\kms\ \cite{sanfrutos2013}. For a black hole mass of $M_{BH} = 1.5\times 10^7$\,$M_\odot$, this is $v_c \leq 0.095$\,$r_{\rm g}$\,ks$^{-1}$. The occultation event characterised in \swi\ lasts $126.4$\,ks, so that considering the upper limit in the velocity as the correct value, the cloud would travel a distance of $12$\,$r_{\rm g}$ within this lapse \cite{sanfrutos2013}. We divide the spectra in 24 equal sub--spectra, consequently each one of these sub--spectra covers $0.5$\,$r_{\rm g}$ in the cloud's orbit. We force the position $\alpha = 0$\,$r_{\rm g}$ to coincide with the lowest flux observations, so that our observation starts when the cloud is at $\alpha = 8$\,$r_{\rm g}$ and ends when it reaches $\alpha = -4$\,$r_{\rm g}$. 

The only parameters that are not frozen are the normalizations of the power law and the reflection component, and the radii of the obscuring cloud and the corona. 
These four parameters are free to vary. The normalization of the reflection component, as well as the cloud and source radii are constrained to be the same for all of the 24 spectra. The normalizations of the power law are allowed to vary independently for every spectrum. 

We performed fits with all the 24 spectra extracted from the \xmm\ observation of \swi . 
The results of the fit are poorly constrained ($\chi^2$/dof\,$\sim$\,$1.63$, see left column of Table\,\ref{tab:swift_fit}). We further investigate any possible improvement by allowing the column density of the cloud to vary. The fit gets much better ($\chi^2$/dof$=1.25$, see right column of Table\,\ref{tab:swift_fit}). 
Absorption is negligible in all spectra above $6-7$\,\kev , becoming important in the soft and intermediate \xr s. 

\begin{table} 
\caption{\label{tab:swift_fit}Fits of GR absorption model to \swi\ data}
\begin{center}
\begin{tabular}{c c c}
\hline\hline
Parameter                        & $N_{\rm H}$ fix        & $N_{\rm H}$ free       \\
\hline
$N_{\rm H} / 10^{22}$\,cm$^{-2}$ & $2^f$                  & $17.5\pm 1.0$          \\
$r^{\rm (cloud)} / r_{\rm g}$    & $4.9\pm 0.2$           & $5.2\pm 0.1$           \\
$r^{\rm (source)} / r_{\rm g}$   & $7.8\pm 0.2$           & $9.0\pm 0.2$           \\
nors. PL                         & $0.8 - 3.4$            & $0.5 - 2.2$            \\ 
nor. REF                         & $1.4\pm 0.2$           & $6.8\pm 0.3$           \\ 
$\chi^2/{\rm dof}$               & $10854/6643 \sim 1.63$ & $8312/6642 \sim 1.25$  \\
\hline
\hline
\multicolumn{3}{c}{
  \begin{minipage}{0.6\textwidth}
Normalizations in units of $10^{-4}$\,photons\,\kev$^{-1}$\,cm$^{-2}$\,s$^{-1}$ at $1$\,\kev . 
All individual PL normalizations are within the range given. 
The superscript `$^f$' means `frozen'. 
  \end{minipage}
}
\end{tabular}
\end{center}
\end{table}

The source, defined as an annulus of inner radius fixed to $3.9$\,$r_{\rm g}$, turns out to have an outer radius of $\left(9.0\pm0.2\right)$\,$r_{\rm g}$. The cloud's radius is $\left(5.2\pm 0.1\right)$\,$r_{\rm g}$, \ie\ the cloud represents around $\sim$\,$50-60\%$ of the source in size ($\sim$\,$40\%$ in projected surface). 
The cloud is only a few times (less than one order of magnitude) thicker than when $N_{\rm H}$ was fixed, with column density of $\left(1.75\pm 0.1\right) \times 10^{23}$\,cm$^{-2}$, in stead of the originally fixed value of $2 \times 10^{22}$\,cm$^{-2}$. 
These results on the geometry of the system are complementary to those obtained when the GR eclipsing cloud code was not used (see \cite{sanfrutos2013}), with cloud and source sizes $D_{\rm c}$\,$\leq$\,$7$\,$r_{\rm g}$, and $D_{\rm s}$\,$\leq$\,$10.5$\,$r_{\rm g}$, respectively: both the cloud and the source are slightly larger, most likely in order to account for the larger column density ($\sim$\,$10^{23}$\,cm$^{-2}$ in stead of $\sim$\,$10^{22}$\,cm$^{-2}$).

\section{Summary}

We have introduced a new relativistic \xr\ spectral model for characterising \xr\ occultation events, involving key parameters such as the size of the \xr\ emitting regions (the innermost accretion disk and\,/\,or the corona), the disk's emissivity index, the column density, ionization state, linear size and velocity of the obscuring cloud, the black hole spin, and the inclination of the system with respect to our LOS. 
Tests on \xmm\ simulated data show that our instruments detect anisotropic emission from the disk: GR effects enhance the emission from the side of the disk approaching towards the observer, while emission from the receding parts decreases. We show how an eclipse of the \xr\ emitting regions by a cloud in the BLR of the system can be used to probe the close environments of the central SMBH.
 
Since the tests carried out have been satisfactory, we fit our relativistic model to real data from a long \xmm\ observation of \swi . 
We define an annulus-like \xr\ source, parallel and close to the accretion disk, whose inner radius is fixed to $3.9$\,$r_{\rm g}$ from previous unabsorbed observations. We compute its outer radius to be $\left(9.0\pm0.2\right)$\,$r_{\rm g}$. The radius of the eclipsing cloud is $\left(5.2\pm 0.1\right)$\,$r_{\rm g}$, and its column density is $\left(1.75\pm 0.1\right) \times 10^{23}$\,cm$^{-2}$. 
As compared to our previous, non-relativistic work, the sizes of the cloud and the source are lightly larger, probably in order to balance the larger column density calculated in the GR scenario.

%
%
\small  
%
\section*{Acknowledgments}   
%
Financial support for this work was provided by the European Union through the Seventh Framework Programme (FP7/2007-2013) under grant n. 312789. 
%

%
\end{document}